\documentclass[journal,comsoc]{IEEEtran}

\usepackage{cite}
\usepackage{amsmath,amssymb,amsfonts,bm}
\usepackage{algorithm}
\usepackage{algorithmic}
\usepackage{graphicx}
\usepackage{epstopdf}
\usepackage{booktabs}
\usepackage{makecell}
\usepackage{multirow}
\usepackage{textcomp}
\usepackage{xcolor}
\usepackage[colorlinks  = true,
            linkcolor   = black,
            urlcolor    = magenta,
            anchorcolor = blue,
            bookmarks   = false
            ]{hyperref}

\def\BibTeX{{\rm B\kern-.05em{\sc i\kern-.025em b}\kern-.08em
    T\kern-.1667em\lower.7ex\hbox{E}\kern-.125emX}}

\newcommand{\overbar}[1]{\mkern 1.5mu\overline{\mkern-1.5mu#1\mkern-1.5mu}\mkern 1.5mu}

\begin{document}

% \title{Towards Free Performance Boosting: Codeword Mimic Learning for Massive MIMO CSI feedback}
\title{Better Lightweight Network for Free: Codeword Mimic Learning for Massive MIMO CSI feedback}

\author{

Zhilin Lu, Xudong Zhang, Rui Zeng and Jintao Wang,~\IEEEmembership{Senior Member,~IEEE}% and Jian Song,~\IEEEmembership{Fellow,~IEEE}

\thanks{
% TODO: any fundation?

The authors are with the Department of Electronic Engineering, Tsinghua University, and Beijing National Research Center for Information Science and Technology (BNRist), Beijing 100084, China. (e-mail: luzl18@mails.tsinghua.edu.cn, zxd22@mails.tsinghua.edu.cn, zengr21@mails.tsinghua.edu.cn, wangjintao@tsinghua.edu.cn).

The key results can be reproduced with the following github link: \textnormal{\href{https://github.com/Kylin9511/CodewordMimicFeedback}{https://github.com/Kylin9511/CodewordMimicFeedback}}.
}% <-this % stops a space
% \thanks{Manuscript received XXX, XX, 2015; revised XXX, XX, 2015.}
}

%\markboth{IEEE Transactions on Vehicular Technology,~Vol.~XX, No.~XX, XXX~2020}
{}
%{Shell \MakeLowercase{\textit{et al.}}: Bare Demo of IEEEtran.cls for Journals}

\maketitle

\begin{abstract}
The channel state information (CSI) needs to be fed back from the user equipment (UE) to the base station (BS) in frequency division duplexing (FDD) multiple-input multiple-output (MIMO) system. Recently, neural networks are widely applied to CSI compressed feedback since the original overhead is too large for the massive MIMO system. Notably, lightweight feedback networks attract special attention due to their practicality of deployment. However, the feedback accuracy is likely to be harmed by the network compression. In this paper, a cost free distillation technique named codeword mimic (CM) is proposed to train better feedback networks with the practical lightweight encoder. A mimic-explore training strategy with a special distillation scheduler is designed to enhance the CM learning. Experiments show that the proposed CM learning outperforms the previous state-of-the-art feedback distillation method, boosting the performance of the lightweight feedback network without any extra inference cost.
\end{abstract}


%%%% Introduction
\section{Introduction}

% 大背景
\IEEEPARstart{M}{assive} multiple-input multiple-output (MIMO) can increase the spectrum and energy efficiency with a larger scale of antennas.
However, it is required for the base station (BS) to obtain the downlink channel state information (CSI) for beamforming to realize its advantages.
In frequency division duplexing (FDD) systems, there is a lack of channel reciprocity between the uplink and downlink CSI.
Therefore, the downlink CSI needs to be first estimated at the user equipment (UE) and then fed back to the BS.
Due to the unacceptable transmitting overhead brought by the huge antenna scale in massive MIMO systems, CSI compressed feedback becomes an important task \cite{he2019modeldriven}.

% 技术发展，性能提升
The deep learning (DL) aided CSI compression has become the mainstream feedback solution since CsiNet \cite{wen2018deep} proved its superiority over the traditional compressed sensing.
Following the pipeline in CsiNet, many influential works are proposed to acquire better feedback performance.
For instance, CsiNet+ \cite{guo2020convolutional} enlarges the receptive field of the convolutional layers to improve the feature extraction ability.
CRNet \cite{lu2020multiresolution} introduces the multi-resolution architecture to capture the features of different scales.
However, many novel feedback networks are heavier than the original CsiNet. A lightweight encoder design is necessary since the hardware resource of a communication system is strictly limited especially for the UE.

% 轻量化设计
In fact, researchers have designed many methods to reduce the network cost.
The fully connected (FC) layer is binarized in \cite{lu2021binary} so that the parameter size is greatly reduced.
ConvCsiNet \cite{cao2021lightweight} introduces ways of compressing the convolutional network.
Additionally, weight pruning is utilized in \cite{guo2020compression} to drop the redundant neural connections. However, network compression is likely to cause performance degradation. It is significant to explore ways of enhancing the lightweight networks.

Knowledge distillation (KD) \cite{hinton2015distilling} is a popular technique to boost the performance of lightweight networks in DL.
It utilizes a powerful but heavier teacher model to guide the training of a lighter student model.
A vanilla KD algorithm is first introduced into CSI feedback in \cite{tang2021knowledge}.
Nevertheless, experiments show that the vanilla KD algorithm does not perform well with a practical extremely lightweight encoder.

% KD
\begin{figure*}[t]
\centering
\includegraphics[width=\textwidth]{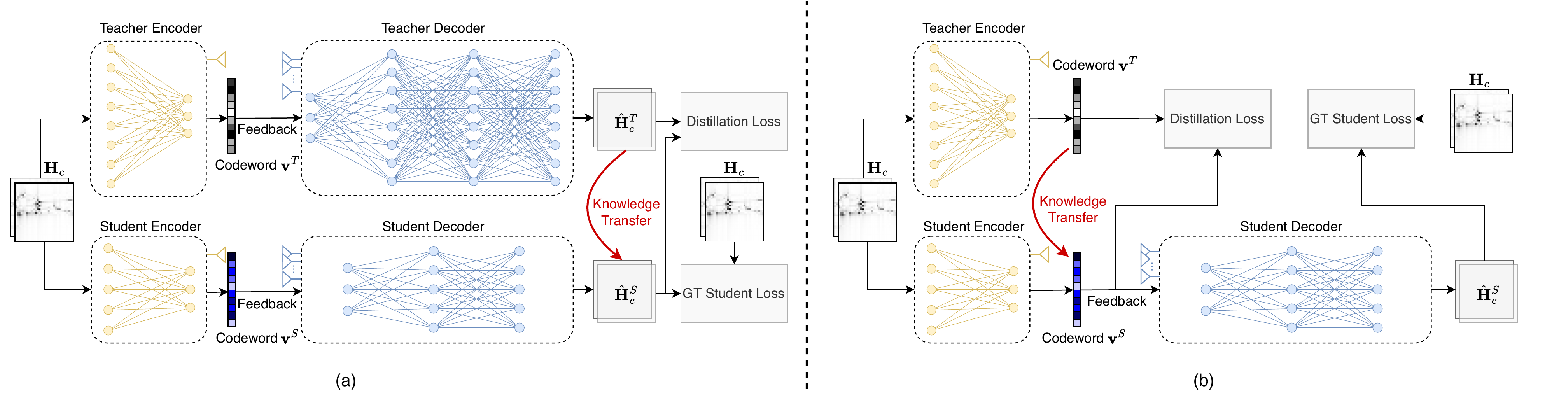}
\caption{The design of the CM learning. (a) presents the previous feedback KD algorithm in \cite{tang2021knowledge}. (b) shows the detailed scheme of the CM aided feedback. As we can see, the transferred knowledge in CM learning is the codeword given by the encoder instead of the reconstructed CSI given by the decoder.}
\label{codeword_mimic}
\end{figure*}

% contribution
In this paper, a new KD scheme is specially designed for the CSI feedback task and proved to be effective for the distillation with the extremely lightweight encoder.
The main contributions of this paper are listed as follows.
\begin{itemize}
    \item A new distillation strategy named codeword mimic (CM) is proposed for lightweight encoders. Codewords generated by the encoder are used as the transferred knowledge. Experiments show that our new scheme outperforms the previous state-of-the-art feedback KD strategy \cite{tang2021knowledge}.
    \item A two-stage training pipeline named mimic-explore is designed for further performance boosting. Experiments show that the CM aided feedback benefits from the balance between the codeword mimic and the CSI recovery.
    \item In order to unlock the full potential of the CM strategy, a cosine annealing distillation scheduler is designed based on ablation studies.
\end{itemize}

%%%% System Model
\section{System Model} \label{Section-SystemModel}

In this paper, a single-cell massive MIMO-FDD system with $N_t$ transmitting antennas at the BS and $N_r$ receiving antennas at the UE is considered.
For simplicity, we have $N_t >> 1$ and $N_r$ is set to $1$.
The orthogonal frequency division multiplexing (OFDM) with $\overbar{N_c}$ sub-carriers is adopted.
The received signal at the $i^{th}$ sub-carrier $y_i$ can be described as follows:
\begin{equation}
\label{system model}
y_{i} = \overline{\mathbf{h}}_i^{H} \mathbf{p}_{i} x_{i} + n_{i},
\end{equation}
where $\overbar{\mathbf{h}}_{i} \in \mathbb{C}^{N_{t} \times 1}$, $\mathbf{p}_{i} \in \mathbb{C}^{N_{t} \times 1}$, $x_{i} \in \mathbb{C}$, and $n_{i} \in \mathbb{C}$ represent the downlink channel vector, beamforming vector, transmitted symbol, and additive Gaussian noise at the $i^{th}$ sub-carrier, respectively.
$(\cdot)^H$ denotes conjugate transpose.

In the FDD system, there is a lack of reciprocity between uplink and downlink channels.
Thus, the UE needs to feed the downlink CSI matrix $\overbar{\mathbf{H}} = \left[ \overbar{\mathbf{h}}_1, ..., \overbar{\mathbf{h}}_{\overbar{N_c}}\right]^H \in \mathbb{C}^{\overbar{N_c} \times N_t} $ back to the BS for beamforming design.
However, the downlink CSI matrix $\overbar{\mathbf{H}}$ contains $2 \overbar{N_c}N_t$ real numbers, which is unacceptably large for the digital feedback.

Considering the channel sparsity in the angular-delay domain, we can first reduce the feedback overhead by transferring the downlink CSI matrix $\overbar{\mathbf{H}}$ into the angular-delay domain with discrete Fourier transform (DFT) as follows:
\begin{equation}
\label{DFT}
\mathbf{H} = \mathbf{A} \overbar{\mathbf{H}} \mathbf{B}^{H},
\end{equation}
where $\mathbf{A} \in \mathbb{C}^{\overbar{N_c}\times\overbar{N_c}}$ and $\mathbf{B} \in \mathbb{C}^{N_t \times N_t}$ are the DFT matrices.
In the angular-delay domain CSI matrix $\mathbf{H}$, only the first $N_c$ rows contain relatively large values while the values in the rest rows are very close to zero. With only the submatrix $\mathbf{H}_c$ fed back, the overhead is reduced to $2 N_c N_t$ real numbers.

\begin{figure}[!b]
\centering
\includegraphics[width=\linewidth]{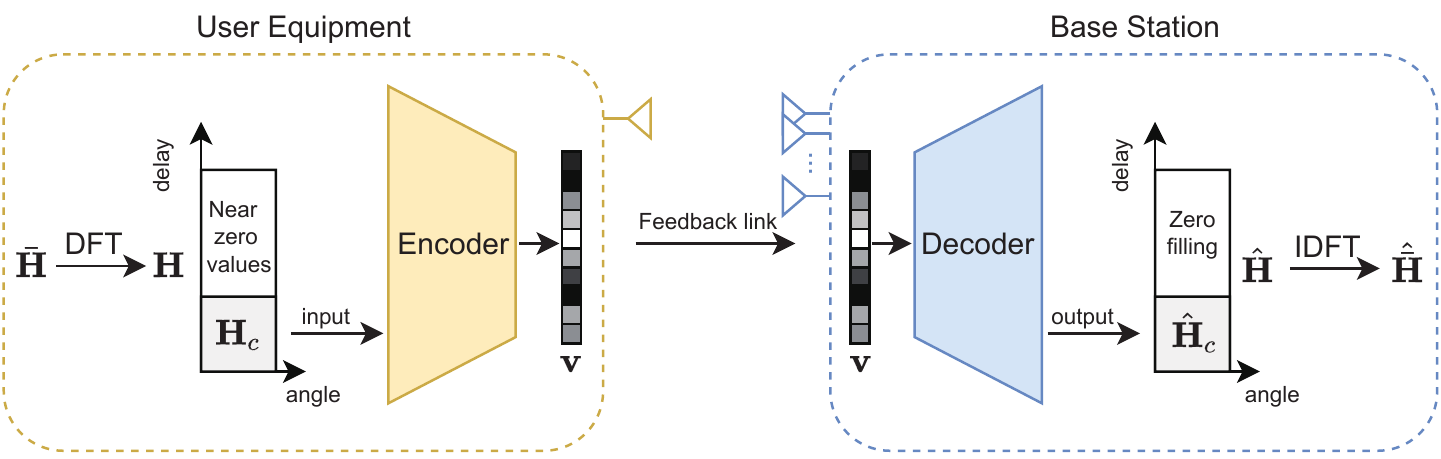}
\caption{The pipeline of the DL-based CSI compressed feedback.}
\label{structure}
\end{figure}

To further reduce the feedback cost, the DL-based CSI compression scheme is utilized.
As shown in Fig. \ref{structure}, the CSI submatrix $\mathbf{H}_c$ is first compressed into a codeword $\mathbf{v} \in \mathbb{R}^{M\times1}$ by the encoder network.
Then, the decoder at the BS recovers the submatrix $\hat{\mathbf{H}}_c$ according to the codeword $\mathbf{v}$ fed back from the UE.
Notably, we define the compression ratio as $\eta=\frac{M}{2 N_c N_t}$.
The whole feedback process can be described as follows:
\begin{equation}
\label{feedback}
\hat{\mathbf{H}}_c = \mathcal{D}\left(\mathcal{E}\left(\mathbf{H}_c, \Theta_\mathcal{E}\right), \Theta_\mathcal{D}\right),
\end{equation}
where $\mathcal{E}\left(\cdot\right)$ and $\mathcal{D}\left(\cdot\right)$ denote the encoding and decoding procedures. $\Theta_\mathcal{E}$ and $\Theta_\mathcal{D}$ represent the parameters of encoder and decoder networks. Once the recovered $\hat{\mathbf{H}}_c$ is obtained, the zero-filling and inverse discrete Fourier transform (IDFT) are conducted to recover the original CSI matrix.

%%%% The proposed Codeword Mimic KD
\section{Codeword Mimic Learning for CSI Feedback} \label{Section-CodewordMimic}

\subsection{Codeword Mimic Based Knowledge Distillation}

In order to train lightweight networks better, vanilla distillation is first introduced to CSI feedback in \cite{tang2021knowledge}. As depicted in Fig. \ref{codeword_mimic}-(a), a heavy but more powerful teacher network takes part in the training and a distillation loss is added to narrow the gap between the output of the teacher $\hat{\mathbf{H}}_c^T$ and that of the student $\hat{\mathbf{H}}_c^S$. However, knowledge transfer between decoder outputs can not guide the training of the encoder well due to the long distance. Considering that the UE is much more resource sensitive, a proper distillation scheme designed for lightweight encoder training is needful.

Instead of transferring knowledge from teacher output to student output, it is better for the lightweight encoder training to directly mimic the codeword produced by the teacher encoder. As we can see in Fig. \ref{codeword_mimic}-(b), our proposed codeword mimic learning is designed to transfer the knowledge from the teacher encoder to the student encoder. The CM strategy introduces a new term to the original MSE loss and the global loss function $\mathcal{L}$ is defined as follows.
\begin{equation} \label{eq-loss}
\begin{aligned}
    \mathcal{L} &= \alpha(t)\mathcal{L}_{cm} + (1-\alpha(t))\mathcal{L}_{gt} \\
    &= \alpha(t)\text{MSE}(\mathbf{v}^T, \mathbf{v}^S) + (1-\alpha(t))\text{MSE}(\mathbf{H}_c, \hat{\mathbf{H}}_c), \\
\end{aligned}
\end{equation}
where $\mathcal{L}_{cm}$ and $\mathcal{L}_{gt}$ are the CM loss and the original ground truth loss, respectively. Codewords given by the teacher and student encoder are denoted as $\mathbf{v}^T$ and $\mathbf{v}^S$. $\alpha(t) \in [0,1]$ is a balance weight of the two losses determined by the current training epoch $t$. The detailed design of $\alpha(t)$ will be discussed in section \ref{SubSection-MimicExplore}.

Compared with previous KD feedback in \cite{tang2021knowledge}, the proposed CM strategy has two advantages. For one thing, the lightweight encoder is more likely to converge towards a better local minima following the guidance of the powerful teacher encoder. This is beneficial when the encoder is compressed to meet the UE deployment constraints. For another, the extra training cost brought by the teacher network is largely reduced since the teacher decoder with dominant complexity is removed.

\subsection{Mimic-Explore Strategy and Distillation Scheduler Design} \label{SubSection-MimicExplore}

In spite of the aforementioned benefits, the CM strategy brings extra challenges to the feedback network training. The main concern is that the codeword mimicking would restrain the network convergence since the optimization is not focused on pure CSI reconstruction. Therefore, a two-stage training strategy named mimic-explore is proposed specially for the CM learning scheme. The main idea is to mimic the codeword only at the early stage. When the student codeword is close enough to the teacher one, the whole network is allowed to explore freely for better CSI reconstruction while keeping the codeword similarity to some extent.

\begin{figure}[!b]
\centering
\includegraphics[width=\linewidth]{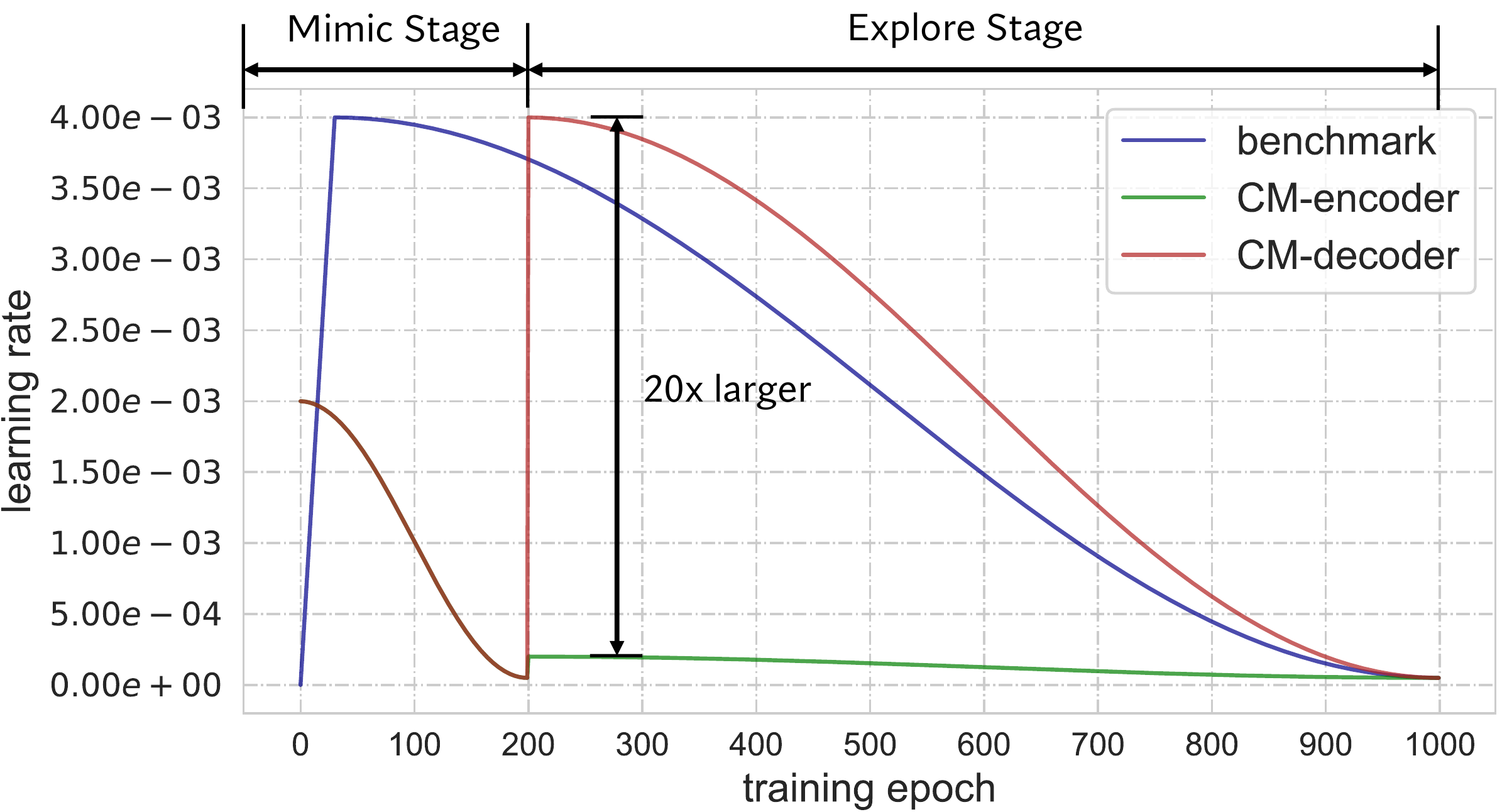}
\caption{The learning rate of benchmark and the mimic-explore strategy.}
\label{mimic-explore}
\end{figure}

The mimic-explore strategy can be easily applied to feedback network training with the learning rate (LR) design. As it is depicted in Fig. \ref{mimic-explore}, the encoder and the decoder share the same cosine annealing LR at the mimic stage, decaying from $2\times 10^{-3}$ to $4\times 10^{-5}$. The loss function of the mimic stage is defined as (\ref{eq-loss}) so that the convergence is guided by codeword mimicking and CSI reconstruction at the same time. After mimicking for hundreds of epochs, the knowledge would be sufficiently transferred from the teacher encoder to the student encoder.

Whereafter, the loss function $\mathcal{L}$ degenerates to the ordinary ground truth loss $\mathcal{L}_{gt}$ and the network is encouraged to explore freely premised on preserving the codeword similarity. Different LR schemes are set for the encoder and the decoder to achieve such intention. As we can see in Fig. \ref{mimic-explore}, the decoder LR decays from $4\times 10^{-3}$ to $4\times 10^{-5}$, which allows a sharp update for the decoder parameters. At the mean while, the initial LR of the encoder is set to $2\times 10^{-4}$, which is so small that the encoder is hardly changed during training. This guarantees the retainment of the knowledge learned at the mimic stage for the CM encoder.

In order to balance the $\mathcal{L}_{cm}$ and $\mathcal{L}_{gt}$ with the best practice at the mimic stage, a cosine annealing distillation scheduler is chosen based on the ablation studies.

\begin{equation} \label{eq-alpha}
  \alpha(t) = \left\{
  \begin{aligned}
      & \frac{1}{2}\alpha_0 \left(1 + \cos\left(\frac{t}{T_{cm}}\pi\right)\right) && t \le T_{cm}\\
      & 0 && t > T_{cm}\;, \\
  \end{aligned}
  \right.
\end{equation}

where $\alpha_0$ is the maximal value of $\alpha(t)$ and $T_{cm}$ is the epochs of the mimic training stage. It is obvious that $\mathcal{L}_{cm}$ is disabled when the current training epoch $t$ reaches the explore stage. Notably, the $\alpha_0$ is set to $10^{-4}$ so that the $\mathcal{L}_{cm}$ and $\mathcal{L}_{gt}$ can be scaled to a similar size.

\subsection{BCRNet: A Typical Lightweight Feedback Network} \label{SubSection-BCRNet}
\begin{figure}[!t]
\centering
\includegraphics[width=\linewidth]{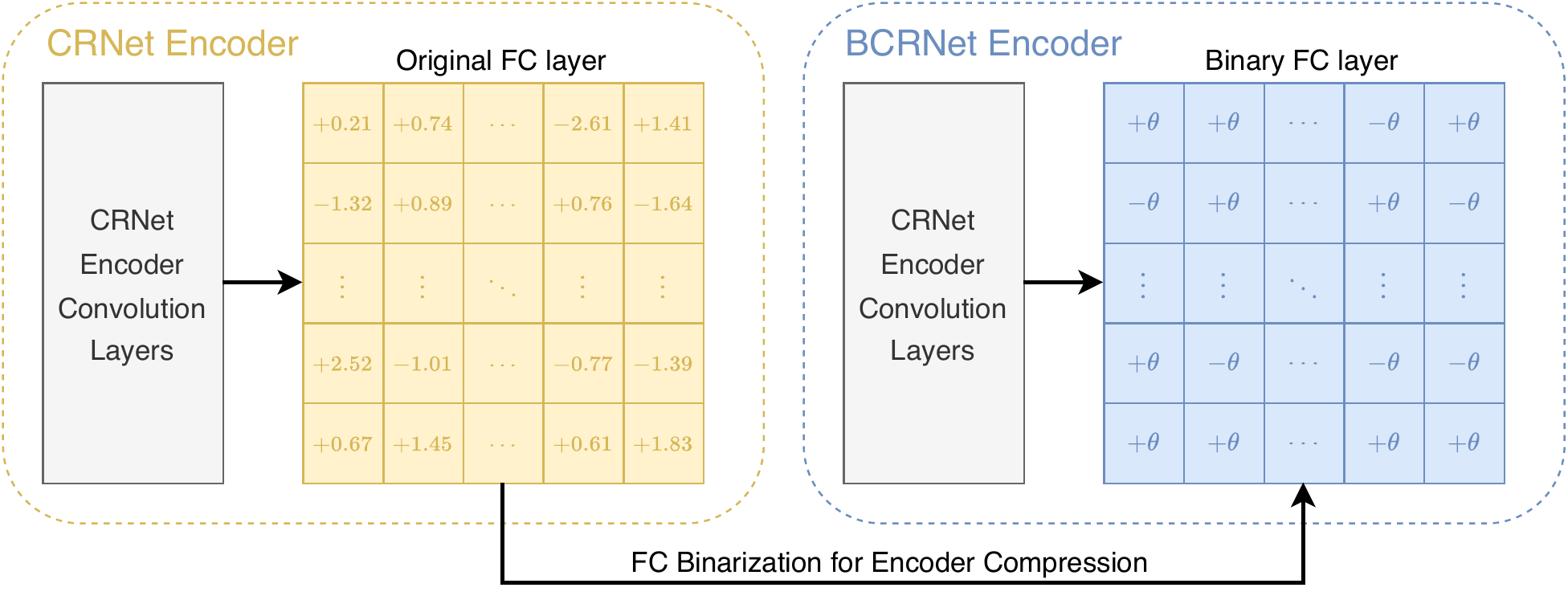}
\caption{The establishment of BCRNet with FC binarization on CRNet encoder. The encoder is largely compressed for practical UE deployment.}
\label{bcrnet}
\end{figure}

The fully connected (FC) layer binarization proposed in \cite{lu2021binary} is a typical feedback encoder compression strategy. As demonstrated in Fig. \ref{bcrnet}, we take the influential CRNet as an example and apply the FC binarization to it, producing binary CRNet (BCRNet) with an extremely lightweight encoder. It is worth mentioning that the decoder of BCRNet is the same as the original CRNet.

The CM learning is a general technique to boost the performance of the feedback network with the lightweight encoder. Since the FC binarization is applicable for a wide range of feedback networks, CRNet and BCRNet can serve as a group of representative teacher and student networks to prove the effectiveness of the proposed CM learning.

%%%% Experiment
\section{Results and Analysis} \label{Section-Experiment}

\subsection{Experimental Settings} \label{SubSection-ExperimentSetting}

The COST2100 dataset \cite{liu2012cost} is adopted following the setting of numerous previous feedback researches including CsiNet, CsiNet+, etc. The training and test datasets consist of 100,000 and 20,000 independently generated samples, respectively. Experiments are carried out under the indoor scenario at 5.3GHz and the outdoor scenario at 300 MHz. $\overbar{N_c}$, $N_c$ and $N_t$ are set to 1024, 32 and 32, respectively.

Adam optimizer is used for the network training under the PyTorch framework. Following the original CRNet, warmup aided cosine annealing LR is adopted for all the benchmark schemes for fair comparison as Fig. \ref{mimic-explore} shows. The batch size is set to 200 and the normalized mean square error (NMSE) is used to evaluate the performance of CSI reconstruction as follows.

\begin{equation}
  \text{NMSE} = \mathbb{E}\left\{ \Vert \hat{\mathbf{H}}_c - \mathbf{H}_c \Vert_2^2 / \Vert \mathbf{H}_c \Vert_2^2 \right\}
\end{equation}

\subsection{Performance of the Proposed Codeword Mimic Learning} \label{SubSection-ExperimentPerformance}

\begin{table}[!t]
\caption{NMSE (dB) and Complexity of the Proposed CM Learning}
\begin{center}
\begin{tabular}{c l|c c | c c}
\Xhline{0.8pt}
\multirow{2}{*}{$\mathbf{\eta}$} & \multicolumn{1}{c|}{\multirow{2}{*}{\textbf{Methods}}} & \multicolumn{2}{c|}{\textbf{Complexity at UE}} & \multicolumn{2}{c}{\textbf{NMSE}} \\
    & & mul$^{\mathrm{a}}$  & params & indoor & outdoor \\
\Xhline{0.8pt}
\multirow{4}{*}{1/4} & \textcolor{gray}{CRNet (teacher)} & \textcolor{gray}{1204K} & \textcolor{gray}{1049K} & \textcolor{gray}{-24.40} & \textcolor{gray}{-11.89} \\
    & BCRNet & 156K & 33K & -17.39 & -8.90 \\
    & BCRNet-KD \cite{tang2021knowledge} & 156K & 33K & -18.26 & -8.82 \\
    & BCRNet-CM & 156K & 33K & \textbf{-19.25} & \textbf{-10.00} \\
\hline
\multirow{4}{*}{1/8} & \textcolor{gray}{CRNet (teacher)} & \textcolor{gray}{680K} & \textcolor{gray}{525K} & \textcolor{gray}{-14.54} & \textcolor{gray}{-8.00} \\
    & BCRNet & 156K & 17K & -13.19 & -6.31 \\
    & BCRNet-KD \cite{tang2021knowledge} &  156K & 17K & -12.72 & -6.47 \\
    & BCRNet-CM & 156K & 17K & \textbf{-13.90} & \textbf{-6.73} \\
\hline
\multirow{4}{*}{1/16} & \textcolor{gray}{CRNet (teacher)} & \textcolor{gray}{418K} & \textcolor{gray}{262K} & \textcolor{gray}{-11.35} & \textcolor{gray}{-5.44} \\
    & BCRNet & 156K & 8K & -8.94 & -4.36 \\
    & BCRNet-KD \cite{tang2021knowledge} & 156K & 8K & -9.69 & -4.19 \\
    & BCRNet-CM & 156K & 8K & \textbf{-10.36} & \textbf{-4.53} \\
\hline
\multirow{4}{*}{1/32} & \textcolor{gray}{CRNet (teacher)} & \textcolor{gray}{287K} & \textcolor{gray}{131K} & \textcolor{gray}{-8.93} & \textcolor{gray}{-3.51} \\
    & BCRNet & 156K & 4K & -7.87 & -2.91 \\
    & BCRNet-KD \cite{tang2021knowledge} & 156K & 4K & -8.12 & -2.71 \\
    & BCRNet-CM & 156K & 4K & \textbf{-8.20} & \textbf{-2.98} \\
\Xhline{0.8pt}
\multicolumn{5}{l}{$^{\mathrm{a}}$ ``mul'' refers to the total number of multiplication.} \\
\end{tabular}
\label{tab1}
\end{center}
\end{table}

\begin{table}[!b]
\renewcommand\tabcolsep{8pt} % adjust the column step of the table
\caption{Performance of Different Mimic-Explore Proportions}
\begin{center}
\begin{tabular}{c | c c | c c}
\Xhline{0.8pt}
\multirow{2}{*}{Mimic$-$Explore} & \multicolumn{2}{c|}{\textbf{Indoor}} & \multicolumn{2}{c}{\textbf{Outdoor}} \\
    & NMSE & $\text{MSE}_{cm}^{end}$ & NMSE & $\text{MSE}_{cm}^{end}$ \\
\Xhline{0.8pt}
  \hspace{0.42cm}$0-1000$ & -17.39 & 27.416 & -8.90 & 371.007 \\
  $100-900$ & -18.85 & 2.058 & -9.93 & 6.041 \\
  $200-800$ & \textbf{-19.25} & 1.977 & \textbf{-10.01} & 6.008 \\
  $300-700$ & -19.01 & 1.544 & -9.82 & 7.192 \\
  $500-500$ & -18.72 & 1.518 & -9.70 & 5.914 \\
  \hspace{-0.42cm}$1000-0$ & -13.90 & \textbf{0.705} & -8.64 & \textbf{4.448} \\
\Xhline{0.8pt}
\multicolumn{5}{l}{``$\text{MSE}_{cm}^{end}$'' refers to the codeword mimic MSE after the explore stage.} \\
\end{tabular}
\label{tab2}
\end{center}
\end{table}

As we can see from Table \ref{tab1}, the student network BCRNet suffers from a prominent performance loss compared with the teacher network CRNet due to the encoder compression. Specifically, the encoder size of the BCRNet is over $30\times$ smaller than the CRNet, which is vastly beneficial to the deployment at UE. And the next challenge is to narrow the performance gap between the BCRNet and the CRNet.

Table \ref{tab1} shows that the NMSE performance of BCRNet is steadily improved with the help of the proposed CM strategy. For instance, the NMSE of BCRNet is decreased for $1.86$dB and $1.11$dB under the indoor and outdoor scenario when the compression ratio $\eta$ is $1/4$. It is worth mentioning that the performance boosting is a result of knowledge transfer from the teacher network and the BCRNet structure is not changed at all. In other words, the improvement is completely cost free for the network deployment.

Moreover, we compare the CM strategy with the previous state-of-the-art distillation strategy \cite{tang2021knowledge}. It can be deduced from Table \ref{tab1} that the vanilla KD strategy is not enough to transfer the knowledge to the student with the extremely lightweight encoder. The proposed CM distillation scheme outperforms the previous KD scheme since the codeword mimic provides more direct guidance for the encoder learning.

\begin{table}[!t]
\renewcommand\tabcolsep{2.8pt} % adjust the column step of the table
\caption{Performance of Different Distillation Schedulers}
\begin{center}
\begin{tabular}{c | c c c | c c c}
\Xhline{0.8pt}
\multirow{2}{*}{$\mathbf{Scheduler}$} & \multicolumn{3}{c|}{\textbf{Indoor}} & \multicolumn{3}{c}{\textbf{Outdoor}} \\
    & NMSE & $\text{MSE}_{cm}^{mid}$ & $\text{MSE}_{cm}^{end}$ & NMSE & $\text{MSE}_{cm}^{mid}$ & $\text{MSE}_{cm}^{end}$ \\
\Xhline{0.8pt}
  const & -17.84 & 0.748 & 2.017 & -9.62 & 4.722 & 6.900 \\
  linear decay & -19.14 & 0.731 & 1.992 & -9.80 & 4.586 & 7.976 \\
  cosine decay & \textbf{-19.25} & \textbf{0.709} & \textbf{1.977} & \textbf{-10.01} & \textbf{4.422} & \textbf{6.008} \\
\Xhline{0.8pt}
\multicolumn{7}{l}{``$\text{MSE}_{cm}^{mid}$'' refers to the codeword mimic MSE after the mimic stage.} \\
\multicolumn{7}{l}{``$\text{MSE}_{cm}^{end}$'' refers to the codeword mimic MSE after the explore stage.} \\
\end{tabular}
\label{tab3}
\end{center}
\end{table}

For realizing the full potential of the proposed CM strategy, ablation studies on the mimic-explore proportion are designed as Table \ref{tab2}. Note that $\eta$ is set to $1/4$ for all experiments and the scheme containing $0$ epochs of mimicking and $1000$ epochs of exploring is the benchmark without the CM learning. It is clear that the final codeword MSE is reduced by dozens of times after adding the CM strategy, proving its effectiveness on the codeword knowledge transfer. As we can see, the targeted NMSE reaches the minimum with $200$ epochs of mimicking. The shorter mimic stage harms the codeword knowledge transfer while the longer mimic stage limits the exploration of the NMSE optimization.

Different distillation schedulers are also tested with $\eta=1/4$ in Table \ref{tab3}. A decaying $\alpha(t)$ gives a better balance for the $\mathcal{L}_{cm}$ and $\mathcal{L}_{gt}$ compared with the simple const $\alpha(t)$. Eventually the cosine annealing $\alpha(t)$ is adopted as equation (\ref{eq-alpha}) since it slightly outperforms the linear decaying scheme.

%%%% Conclusion
\section{Conclusion} \label{Section-Conclusion}

In this paper, a distillation strategy named codeword mimic (CM) was specially designed for the CSI feedback task. With the knowledge transferred from the teacher's codeword, the feedback network with extremely lightweight encoder could achieve better performance without any extra inference cost. In addition, the special mimic-explore strategy and the distillation scheduler were designed to boost the performance of the CM learning. Experiments showed that the proposed CM learning outperformed the previous state-of-the-art feedback distillation scheme and improved the CSI reconstruction quality of the lightweight network under different compression ratios.

\ifCLASSOPTIONcaptionsoff
  \newpage
\fi

% Reference
\bibliographystyle{IEEEtran}
\bibliography{CMKDFeedback.bib}

% Biography
% \begin{IEEEbiography}{Yuguang ``Michael'' Fang}
% Biography text here.
% \end{IEEEbiography}

%It is not necessary to upload the biography when you submit your manuscript.

\end{document}